\documentclass[acmsmall]{acmart}

\usepackage{multirow}
\usepackage{xcolor} 
\definecolor{change}{rgb}{0,0,0}
\definecolor{contribution}{rgb}{0,0,0}

\AtBeginDocument{%
  \providecommand\BibTeX{{%
    \normalfont B\kern-0.5em{\scshape i\kern-0.25em b}\kern-0.8em\TeX}}}

\setcopyright{rightsretained} 
\acmJournal{PACMHCI}
\acmYear{2021} \acmVolume{5} \acmNumber{CSCW2} \acmArticle{479} \acmMonth{10} \acmPrice{}\acmDOI{10.1145/3479866}

\received{April 2021}
\received[accepted]{July 2021} 

\begin{document}

\title{Thing Constellation Visualizer: Exploring Emergent Relationships of Everyday Objects}


\author{Yi-Ching (Janet) Huang}
\affiliation{%
  \institution{Eindhoven University of Technology}
  \country{The Netherlands}}
\email{y.c.huang@tue.nl}

\author{Yu-Ting Cheng}
\affiliation{%
  \institution{Eindhoven University of Technology, The Netherlands \& National Taiwan University of Science and Technology, Taiwan}}
\email{y.cheng@tue.nl}

\author{Rung-Huei Liang}
\affiliation{%
 \institution{National Taiwan University of Science and Technology}
 \country{Taiwan}}
\email{liang@mail.ntust.edu.tw}

\author{Jane Yung-jen Hsu}
\affiliation{%
  \institution{National Taiwan University}
  \country{Taiwan}}
\email{yjhsu@csie.ntu.edu.tw}

\author{Lin-Lin Chen}
\affiliation{%
  \institution{Eindhoven University of Technology}
  \country{The Netherlands}}
\email{l.chen@tue.nl}

\renewcommand{\shortauthors}{Yi-Ching (Janet) Huang et al.}

\begin{abstract}
Designing future IoT ecosystems requires new approaches and perspectives to understand everyday practices. While researchers recognize the importance of understanding social aspects of everyday objects, limited studies have explored the possibilities of combining data-driven patterns with human interpretations to investigate emergent relationships among objects. This work presents \textit{Thing Constellation Visualizer (thingCV)}, a novel interactive tool for visualizing the social network of objects based on their co-occurrence as computed from a large collection of photos. ThingCV enables perspective-changing design explorations over the network of objects with scalable links. Two exploratory workshops were conducted to investigate how designers navigate and make sense of a network of objects through thingCV. The results of eight participants showed that designers were actively engaged in identifying interesting objects and their associated clusters of related objects. The designers projected social qualities onto the identified objects and their communities. Furthermore, the designers changed their perspectives to revisit familiar contexts and to generate new insights through the exploration process. This work contributes a novel approach to combining data-driven models with designerly interpretations of thing constellation towards More-Than Human-Centred Design of IoT ecosystems.
\end{abstract}




\maketitle

\section{Introduction}
\label{sec:intro}
Internet of Things (IoT) has envisioned that every object can be connected and distributed ubiquitously to perform tasks autonomously to fulfil people's needs~\cite{Ashton:1999,Atzori:ComputerNetworks2010}. While IoT research has largely focused on designing new devices, {\color{change}limited studies explore ways to appropriately integrate new devices into our everyday practice. In everyday practice, there are already various objects designed or used for supporting our needs and daily activities. To approach the IoT vision where objects can cooperate with each other to reach common goals, researchers need to investigate how people interact with existing objects in practice~\cite{Giaccardi:Interaction2015}. With an understanding of existing relationships between people and objects, designers can design a better IoT ecosystem and integrate it meaningfully into our everyday life. As such,} recent researchers have emphasized the need of looking beyond what has currently been made in IoT design, particularly exploring new ways to understanding social relationships between humans and everyday objects~\cite{Crabtree:CSCW2016,Williams:CHI2020}. However, there are still limited approaches and tools for exploring these new design possibilities. This work aims to address these challenges to develop a novel tool for supporting designers to understand the social relationships among mundane objects in everyday life.

Researchers have investigated various ways to understand relationships between humans and objects by changing the ways of perceiving the world~\cite{Giaccardi:DIS2016,Giaccardi_Cila:2016,Chang:DIS2017}. For example, Giaccardi et al. propose using the ``thing perspective'' to understand complex relationships between humans and objects through a non-human view~\cite{Giaccardi:DIS2016,Giaccardi_Cila:2016}. Wakkary et al. use connected things (e.g., bowls and cups) to rethink social relationships between people and objects, highlighting objects' parallel social lives that people do not need to be involved in~\cite{Wakkary:DIS2017}. While these unique perspectives are valuable for future IoT design, the great challenges of understanding emergent relationships among objects from everyday practices still remain. First, such emergent relationship is hidden and barely recognized by people. It requires experienced designers or experts to extract these patterns from empirical data through an iterative sense-making process. Second, as the volume and complexity of data increases, it becomes more challenging even for experts to extract these patterns if they don't have appropriate support. In addition, these observations around a single object or a few objects are still limited due to the lack of tools.

To move beyond a limited perspective around objects, some researchers have articulated the importance and challenges of constellation design. {\color{change} Researchers have used constellation as a metaphor to describe the complex relationships where people, objects, environments, and data are entangled~\cite{Coulton_Lindley:2019, Murray-Rust:HTTF2019}.} While recent studies mainly focus on a theory-driven exploration to provide a conceptual or theoretical understanding of constellations~\cite{Frauenberger:TOCHI2019,Murray-Rust:HTTF2019,Lindley:2020}, {\color{change}there is a lack of studies exploring a practical approach and tool to support designers to investigate constellation design in everyday practice. Therefore,} researchers further highlight the urgent need of developing new techniques that allow designers to work collaboratively with computer algorithms to make sense of empirical data~\cite{Murray-Rust:HTTF2019}. 

To address this challenge, this work takes the first step to develop a tool with a computational approach to enable designers to make sense of emergent relationships of everyday objects in real-world practice {\color{contribution}for supporting IoT ecosystem design. To support human creativity and innovation, the CSCW community has explored various tools or strategies to facilitate an individual's or a group's ideation process by leveraging data and the help of computer algorithms~\cite{Frich:CHI2019,Shi:CSCW2017,Koch:CSCW2020,Wang:CSCW2011}. While data contain rich events and interesting patterns as good stimuli for the ideation process, researchers argue that data need further processing and need humans to intervene, interact and interpret to make them meaningful~\cite{Gaver:CHI2007,Dourish:2018}. Therefore, this work focuses on exploring an appropriate approach that integrates data-driven patterns with human interpretations into the tool design.} In particular, we explore two research questions, with the aim of understanding what new knowledge or unique perspectives we can obtain by using social networks of things as a design material: (1) How can we design a tool that captures emergent relations of everyday objects from real-world data? (2) How do designers use our designed tool to make sense of and interpret a social network of things?

\begin{figure}[h!]
    \includegraphics[width=\textwidth]{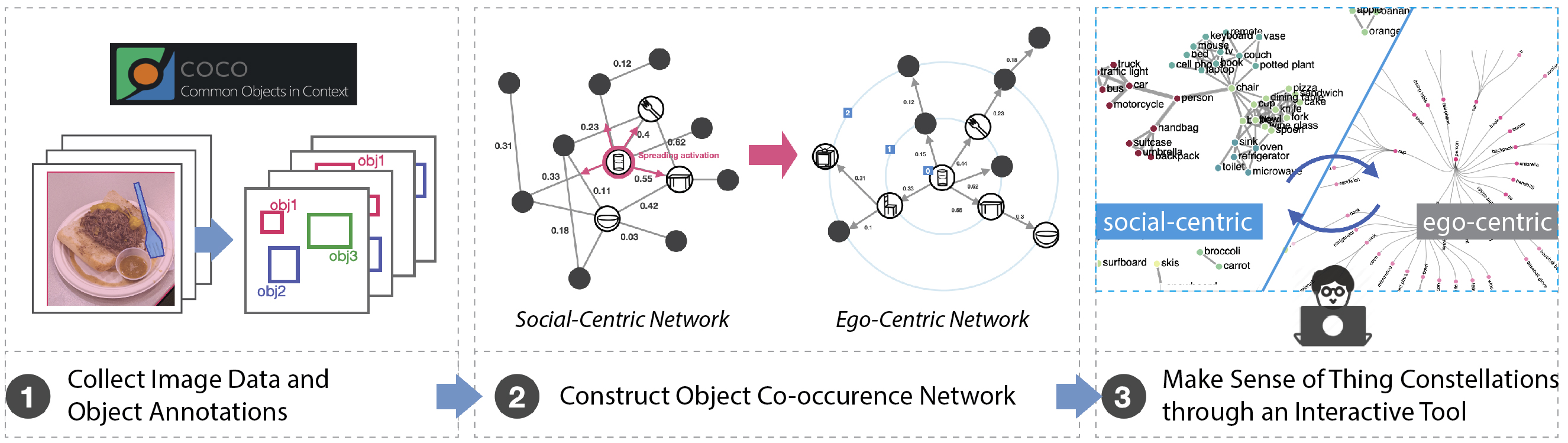}
    \caption{{\color{change}This work presents a novel approach that enables designers to explore a social network of everyday objects through a data-driven exploratory workflow: 1) collect images and annotations, 2) construct two types of object-occurrence networks, and 3) make sense of Thing Constellation through an interactive tool.}}
    \label{fig:data-driven-workflow}
\end{figure}

We introduce \textit{Thing Constellation Visualizer (thingCV)}, an interactive tool that allows design researchers to explore a co-occurrence social network of everyday things ({\color{change}see Figure~\ref{fig:data-driven-workflow}}). With this tool, we propose a computational approach that constructs a social network among things (e.g., human, animals, objects) based on image data which consist of everyday objects across various contexts. To enable flexible design explorations, we provide two views, global and local, to allow designers to make sense of thing constellations from different angles. For the global view, we construct a social-centric constellation among objects by calculating the co-occurrence of every pair of objects in an image based on a large-scale dataset. For the local view, we present an ego-centric constellation by expanding a specific object as a focus node towards an ego-centric social network. 

To further visualize relationships at different levels/granularity, we design an interactive tool for supporting designers to explore the social relationships among objects by using a threshold slider and a perspective-switch button. This tool can support designers to observe the changing characteristics of object communities to understand how groups of things emerge from a single node to a group of communities in an interactive way. Also, designers can flexibly switch between two constellations: zooming in to an ego-centric constellation to see the details and zooming out to a social-centric constellation to see the global structure. 

We conducted two exploratory workshops to investigate how designers use our tool (thingCV) to make sense of a network of objects through a data-driven exploratory process. In the workshops, Eight designers from industry and academia explored a social network of objects which were constructed from the MS-COCO dataset~\cite{Lin:ECCV2014}, which contains everyday objects in diverse contexts. The constructed network of everyday objects has 80 nodes and 2,686 weighted edges, which was built on 123K images. 

The results from the two workshops showed that designers were engaged in identifying interesting (e.g., connected or, conversely, isolated) objects and projecting social quality (e.g., sociable or lonely) onto the identified objects. They also discovered or generated diverse contexts (e.g., familiar but unnoticed contexts or hidden contexts) by identifying object communities. Interestingly, they were able to change their perspectives to revisit familiar everyday practices and generate new insights through this process. We discuss the benefits of combining a computational approach with designerly interpretation in understanding constellations among things and how it can contribute an inspiring perspective to future IoT constellation design.

The paper presents thingCV that makes the following contributions:
\begin{itemize}
    \item A computational approach to constructing a social network of objects based on co-occurrence as computed from a large collection of photos.
    \item An interactive visualization tool that enables perspective-changing design exploration by allowing users to flexibly switch between a social-centric thing constellation and an ego-centric thing constellation. 
    \item Insights from two exploratory workshops showed that thingCV successfully facilitated designers to project social qualities onto the network of objects, to discover alternative contexts, and to change their own perspectives from human-centred to more-than human-centred perspectives.
\end{itemize}

\section{Related Work}
\label{sec:related-work}
\subsection{The Role of Domestic Things for IoT design}
The vision of the Internet of Things (IoT) suggests that computing can be embedded in anything, even in the most mundane objects (e.g., cup, fork, bottle, etc)~\cite{Ashton:1999}. While IoT research has largely focused on designing new devices, exploring new interactions between humans and objects could provide insights and inspire future IoT design. Recent studies have investigated everyday objects at home to understand daily mundane routines between family and objects~\cite{Crabtree:CSCW2016} and rethink possibilities of reconfiguring objects to meet the household's specific needs~\cite{Williams:CHI2020}. Crabtree and Tolmie offer empirical insights of a day in a life of things in the home and unpack distinct categories of everyday things and social patterns of human-thing interaction~\cite{Crabtree:CSCW2016}. Williams et al. contribute empirical insights about how family members imagine a future smart home that incorporates their existing everyday objects by reconfiguring an object's role in the home and rethinking evolving relationships of humans and objects~\cite{Williams:CHI2020}. In this paper, we aim to explore the emergent relationships between people and objects, but from a different angle, by making use of object co-occurrence patterns and human interpretations together. 

\subsection{From Things to Social Things}
Giaccardi et al. proposed Thing Ethnography that allows everyday objects (e.g., kettles and mugs) equipped with sensors and cameras to capture social practices and the patterns of use on a daily basis~\cite{Giaccardi:DIS2016,Giaccardi_Cila:2016}. The aim was to understand the evolving use and applications of things in everyday life by introducing a new object-centred perspective to perceive the world. They suggested that thing’s perspectives could help designers discover unexpected and invisible relationships among objects from unique perspectives that could not be discovered through (human-centred) observations and interviews~\cite{Giaccardi:DIS2016}. In another work, Chang et al. applied a thing’s perspective to smart mobility design, by equipping a motorcycle with cameras to understand the ``life'' of a motorcycle in a specific cultural context from its perspective~\cite{Chang:DIS2017}. Cheng et al. built a camera to capture the relationship among multiple things~\cite{Cheng:DIS2019}.  

Envisioning a day when everyday objects and systems can access the Internet to share data and interact with each other, some researchers have developed research prototypes to understand the complex relationship between things and people for IoT design~\cite{Wakkary:DIS2017,Nicenboim:2018}. One representative example is Morse Things~\cite{Wakkary:DIS2017}. Morse Things are sets of ceramic bowls and cups that can communicate with each other over a home's network. With the connection capability, Morse Things not only communicate with other things but also their human roommates. These things can send dots and dash as Morse codes to each other to know whether the other thing is there. While the bowls and cups can be used to eat or drink by humans, they could still have their ``social life'' with other things that do not need to be shared with their human owners. Morse Things extends the thing ethnography~\cite{Giaccardi:DIS2016} towards different directions and forces people to rethink the social relationships between things and people in a broader networked perspective.  

\subsection{Entangled Ethnography and Constellation}
Another line of research seeks to expand thing-centred design towards a broader view and emphasize the needs of constellation design where people, objects, data, algorithms, and the environment are entangled~\cite{Frauenberger:TOCHI2019, Murray-Rust:HTTF2019, Coulton_Lindley:2019}. Recent work puts much effort into shaping the notion of constellations through a theory-driven exploration~\cite{Frauenberger:TOCHI2019,Murray-Rust:HTTF2019,Coulton_Lindley:2019}. Coulton and Lindley used a constellation metaphor to encompass the interdependent and independent relationships between humans, non-human actants, and environments. They demonstrated how the design could be put into practice through a speculative design~\cite{Coulton_Lindley:2019}. Furthermore, Murray-Rust and Burnett developed Entangled Ethnography that uses a theory-driven perspective to illustrate the vision in which humans, objects, and data are entangled and the computational intelligence will also be used to collaboratively make sense of data with researchers and networks of people and objects~\cite{Murray-Rust:HTTF2019}. Frauenberger further develops Entanglement HCI grounded on entanglement theories to reframe knowledge production practices in HCI, focusing on the performative relationship between humans and technology and ethical challenges~\cite{Frauenberger:TOCHI2019}. The common goal of these studies is to push forward the design paradigm that moves from human-centred toward more-than human-centred design.

\subsection{The Role of Co-occurrence in Investigating Social Relationships of Things}
Recent studies often capture object co-occurrence to investigate the social relationships of things in practice. Object co-occurrence, the pattern of two or more things present in the same place, allows design researchers to capture the practical aspect of the changing relationships in the mundane everyday. Such co-occurrence is not only easily captured by the tool (e.g. camera) but also provides an open space for people to interpret the possible meanings related to social relationships. For example, Giaccardi et al.~\cite{Giaccardi:DIS2016} captured the pattern of any things (people, objects) that are nearby or in front of the object and interpreted the thing-centred relationships in practice. Cheng~\cite{Cheng:DIS2019} also captured object co-occurrence patterns in a home and discussed the meaning behind such co-occurrence with the participants. Desjadrins et al.~\cite{Desjardins:CHI2020} took a step forward by not only capturing the patterns but also speculating about the possible data stream between them, as a way to inspire people to envision the future IoT data design. In these examples, object co-occurrence enables complex social relationships to be investigated with limited data input. The results are also easily observed and understood by people. Therefore, we propose object co-occurrence as a potential anchor to stimulate people’s curiosity to interpret the possible contexts and the meanings around everyday things. 

{\color{contribution}
\subsection{IoT as Cooperative Work between Objects}
While the IoT research has envisioned that objects, people and spaces will be intertwined and rely on some kind of cooperation to achieve common goals, we cannot ignore the socio-technical aspects between objects, people, and spaces. Atzori et al. emphasize that large numbers of objects are ``able to interact with each other and cooperate with their neighbors to reach common goals~\cite{Atzori:ComputerNetworks2010}.'' These activities may or may not be directly linked to cooperative work between people. Therefore, we need to consider the cooperation between objects more seriously and regard it as a kind of cooperative work. This is an important perspective informed by insights of CSCW. Although we learned from Actor Network Theory (ANT) the value of taking the non-human actor into account in socio-technical networks, the notion of objects and other non-human actors cooperating with each other to achieve common goals is not relatively new in CSCW~\cite{Robertson:ECSCW2015}. However, while prior work focuses on exploring theory-driven conceptual frameworks, our work aims to bridge the gap between conceptual, theoretical visions and design practice. The insights of possible interactions and cooperations among objects can contribute to the CSCW community and bring new social-technical aspects into IoT cooperative scenarios.

\subsection{Computer-Supported Creativity in CSCW}
The HCI and CSCW community have established a great interest in understanding various approaches and tools to facilitate human innovation and creativity, including individual or group creativity~\cite{Frich:CHI2019,Shi:CSCW2017,Koch:CSCW2020,Wang:CSCW2011}. Prior work has used computational visual stimuli to facilitate group creativity~\cite{Shi:CSCW2017} and explored factors such as group dynamics in enhancing the ideation process~\cite{Wang:CSCW2011}. Researchers found that appropriately incorporating intelligent tools and algorithms can support diverse human-computer partnerships on collaborative ideation~\cite{Koch:CSCW2020}. While these studies contribute to various strategies to the general ideation process, our work focuses more on the unique challenge of IoT ecosystem design. We extend the prior literature on computer-supported creativity and contribute a novel tool that enables designers to play with social patterns extracted from large-scale empirical data. Particularly, this tool aims to facilitate a group of designers to discover familiar but unnoticed experiences and generate diverse future IoT scenarios.

The strength of our tool is to allow people to explore data-driven patterns flexibly and add their own interpretations upon data. Align with what Gaver et al. suggested~\cite{Gaver:CHI2007}, the potential benefits of combining automatic inferencing and ambiguous output is to encourage user interpretation. They emphasize that human interpretation is needed to make data more meaningful. Dourish and G\'{o}mez Cruz also argue that data ``do not speak for themselves'' and that narratives are necessary to contextualize in order to give them meaning and shape~\cite{Dourish:2018}. Our work demonstrates the great potential of combining data-driven patterns with human interpretations and the ability of the tool to enable people to discover more interesting everyday phenomenons. The new insights generated from the exploratory process will contribute to IoT ecosystem design.
}

\section{Design Goals for Visualizing Thing Constellation}
\label{sec:thing-constellation}
`Constellation' has been proposed by prior design research as a design metaphor to describe any objects in the IoT that ``exist individually, their meaning and significance are augmented by virtue of being part of a wider constellation''~\cite{Lindley:2020}. Based on this metaphor, we use \textit{thing} as a generic term to indicate living and nonliving objects as nodes in the network. We use \textit{constellation} to describe a collective of things that can be linked to form a pattern, and that can have a meaning. Thing Constellation is a virtual artifact shaped by everyday things; it projects the social relationships between objects and forms a possible collaboration or communication enabled by IoT. Thus, it could bring more stories and meanings beyond a single object. To enable designers to explore social links among everyday objects, we identify the following design goals for our tool.

\subsection{Use Co-occurrence as a Starting Point for Exploring Social Links among Objects}
To shape Thing Constellation, the fundamental step is to define the meaning of the `social links' among objects. A link can be defined in different ways~\cite{Atzori:ComputerNetwork2012,Wuest:2012,Giaccardi:DIS2016}, depending on different focuses. Giaccardi et al. identified social links based on the use frequencies of objects~\cite{Giaccardi:DIS2016,Wakkary:DIS2017}. The focus is to explore not only the relationships between human owners and objects but also objects' own relationships with other objects. Nansen et al. emphasized that an object can be not only a social actor but also an object user~\cite{Nansen:OzCHI2014}. 

In this work, we aim to construct a social network of objects from real-world data for designers, so they are able to actually play with networked objects digitally. Thus, practicality and implementation are very important considerations. We use object co-occurrence as a starting point for exploring social links among objects. The resulting co-occurrence social networks are a basic form of social networks, which is commonly used to capture potential relationships between people, concepts, or other entities. In addition, the co-occurrence technique is easily implementable, so it is an appropriate first step to explore the possibility of Thing Constellations.

\subsection{Look for a Large-Scale Dataset Containing Objects in Diverse Contexts}
To visualize an object co-occurrence social network, we need a large amount of data to capture general emergent patterns in practice. Prior research focuses on collecting images or sensor data in the wild. The common challenge is around scaling up and privacy issues. By considering the amount of data and privacy concerns, we decided to look for a large-scale dataset that contains everyday objects in a variety of contexts that are voluntarily contributed by the general public.

\subsection{Provide an Interactive Tool to Enable Flexible Sensemaking on Data}
To support flexible sense-making, we aim to provide an interactive tool that presents diverse perspectives to explore a social network of objects. The interface should be flexible and should enable people to manipulate parameters to observe, to interpret, and to reflect about object co-occurrences and their own personal experiences.

\subsection{Provide Data-Driven Patterns for Encouraging Open Discussions}
To facilitate designers to interpret everyday alternatively, we only present the computational patterns driven by data and provide an interactive interface for people to contribute their subjective interpretations. By doing so, designers will not be influenced by others’ biases but can make interpretations freely. Designers will also not be dominated by other people but share their intuitive thoughts and explore ideas based on the facts.

\section{Thing Constellation Visualizer: Visualizing Thing Constellation through Objects Co-occurrence}
\label{sec:tool-design}

We propose a computational approach that constructs a social network among objects based on objects' co-occurrence in a large-scale image dataset. The overall data-driven exploratory workflow consists of three steps: 1) collect image data and object annotations, 2) construct two types of object co-occurrence networks, and 3) make sense of Thing Constellations by an interactive tool (see Figure~\ref{fig:data-driven-workflow}). 

In this section, we introduce the core technology to construct Thing Constellations from image data and object annotations and visualize them from two different perspectives. There are two types of constellations: social-centric constellation and ego-centric constellation. The goal of the former is to show the global picture of the entire network; the latter one is to present the ego view by expanding a personal social network under a constraint condition.

\subsection{Social-Centric Constellation}
\begin{figure}[h!]
    \includegraphics[width=\textwidth]{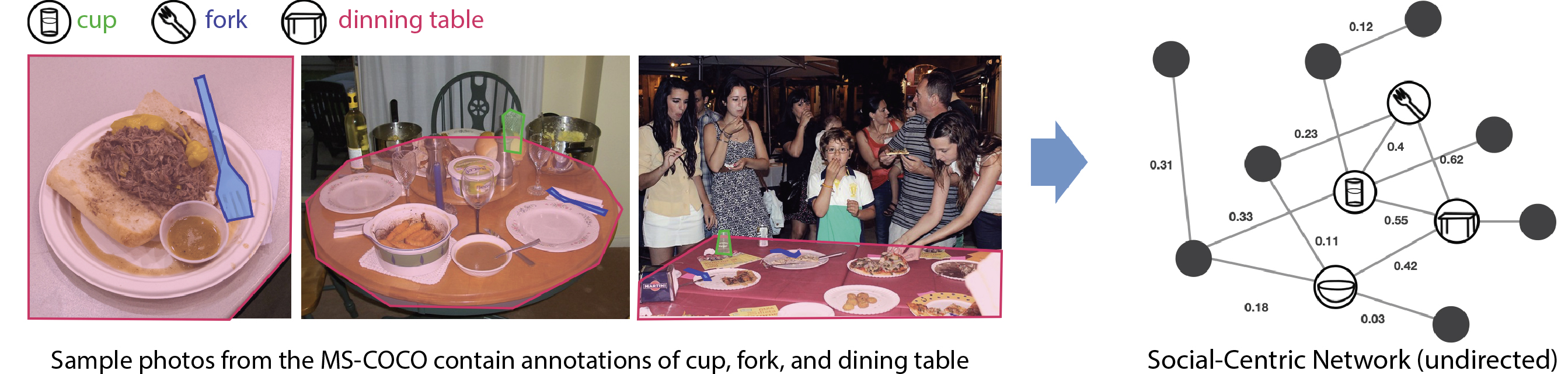}
    \caption{{\color{change}Social-centric network is a network of objects based on their co-occurrence in the same images.}}
    \label{fig:sample-photos-MSCOCO}
\end{figure}


We build a network of objects based on their co-occurrence in the same images. The main assumption is that two objects are more likely to have a close relationship if they frequently appear or be used together. For example, forks are more likely to co-occur with a dining table or a bottle because they are frequently used by people while eating food (see Figure~\ref{fig:sample-photos-MSCOCO}). We use \textit{Jaccard similarity} to measure asymmetric relative co-occurrence between objects. Let $A$ and $B$ be the sets of images containing $object_{A}$ and $object_{B}$, respectively. The relative co-occurrence is defined as the size of the intersection of set $A$ and set $B$ (i.e. the number of common images) over the size of the union of set $A$ and set $B$ (i.e. the number of unique images) (see Equation~\ref{eq:similarity}). $\textit{Relation}(A, B)$ is the relative co-occurrence of $A$ and $B$. $|A \cap B|$ is the number of images in which two objects co-occur and $|A \cup B|$ is the number of photos in which appear in any one of the two objects. In other words, we compute the proportion of object overlapping as object similarity. In the end, we can build the entire network structure by calculating the similarity of any two nodes and assign the similarity score to every link.

\begin{equation} 
\label{eq:similarity}
\textit{Relation}(A,B) = \frac{|A \cap B|}{|A \cup B|}
\end{equation}

\subsection{Ego-Centric Constellation}
\begin{figure}[h!]
    \includegraphics[width=\textwidth]{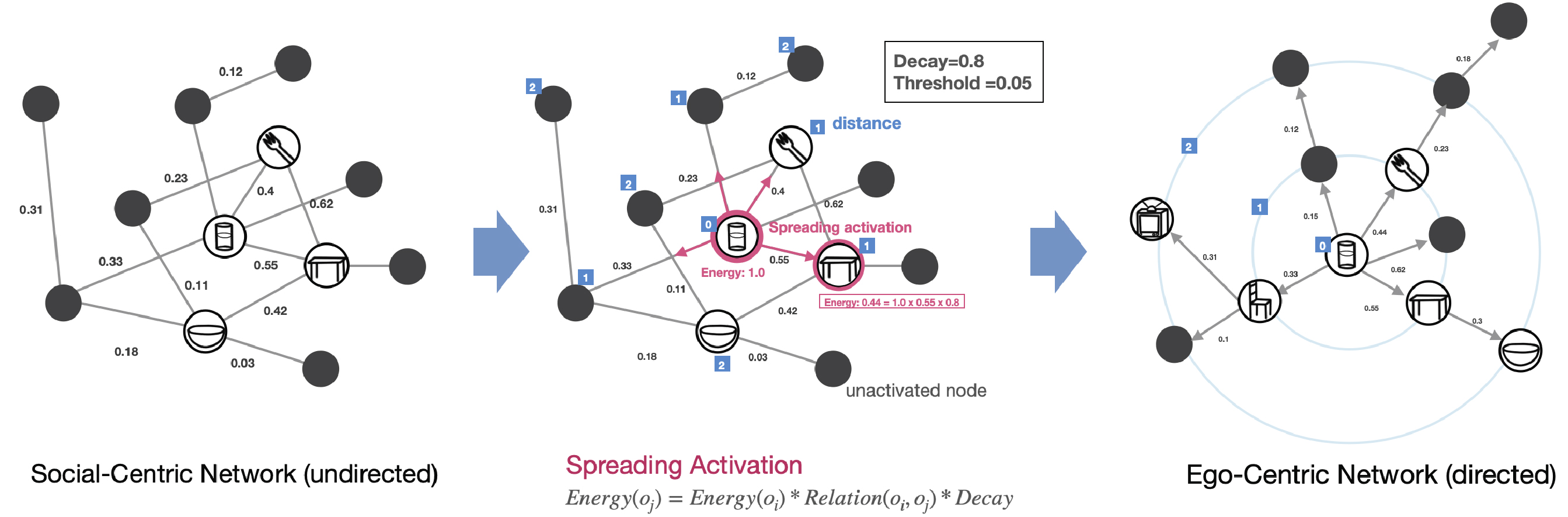}
    \caption{{\color{change}Ego-centric network is retrieved in the social-centric network by spreading energy from a target node to neighboring nodes.}}
    \label{fig:spreading-activition}
\end{figure}


To understand a social structure of a specific object, we use the spreading activation technique to expand an ego-centric network~\cite{Crestani:1997}. Spreading activation is a method of searching associative networks. The process is initiated by giving an activation value (i.e., $E=1.0$) to a single node, and then the value is propagated through the network by gradually decaying the value (e.g., $d = 0.8$) till below a threshold (e.g., $t=0.05$). It is similar to the breath-first-traversal of the graph where the activation is spread to all neighbour nodes, but our method only spreads activation to nodes where they have a link with a weight that is higher than the threshold (see Figure~\ref{fig:spreading-activition}). The energy for each node is defined in Equation~\ref{eq:energy-spreading}. The energy is propagated to neighbouring nodes only if the remaining energy is above the threshold. 
\begin{equation} 
\label{eq:energy-spreading}
\textit{Energy}(o_j) = \textit{Energy}(o_i) * \textit{Relation}(o_i, o_j) * \textit{Decay}
\end{equation}

\subsubsection{Community detection}
To understand the network structure, we use the \textit{Louvain} algorithm to detect communities~\cite{Blondel:2008}. The Louvain algorithm is a hierarchical clustering method that recursively merges communities into a single node and executes the modularity clustering on the condensed graphs. This algorithm separates the network in communities by optimizing the modularity (i.e., the structure of the network) after trying various grouping operations on the network. This algorithm is very computationally efficient for detecting communities in a large and complex network.

\subsection{Dataset}
\label{subsec:dataset}
To validate our idea, we decide to choose an existing open-source dataset rather than a self-collected dataset because we want to focus on designing rich interactive experiences and interactions for our tool rather than spend too much time on the early stage of the data collection process. We are also aware of the potential ethical issues of using open source data as research exploration; the biases might cause some issues. We will discuss these concerns at the end of this paper.

\begin{figure}[h!]
    \includegraphics[width=\textwidth]{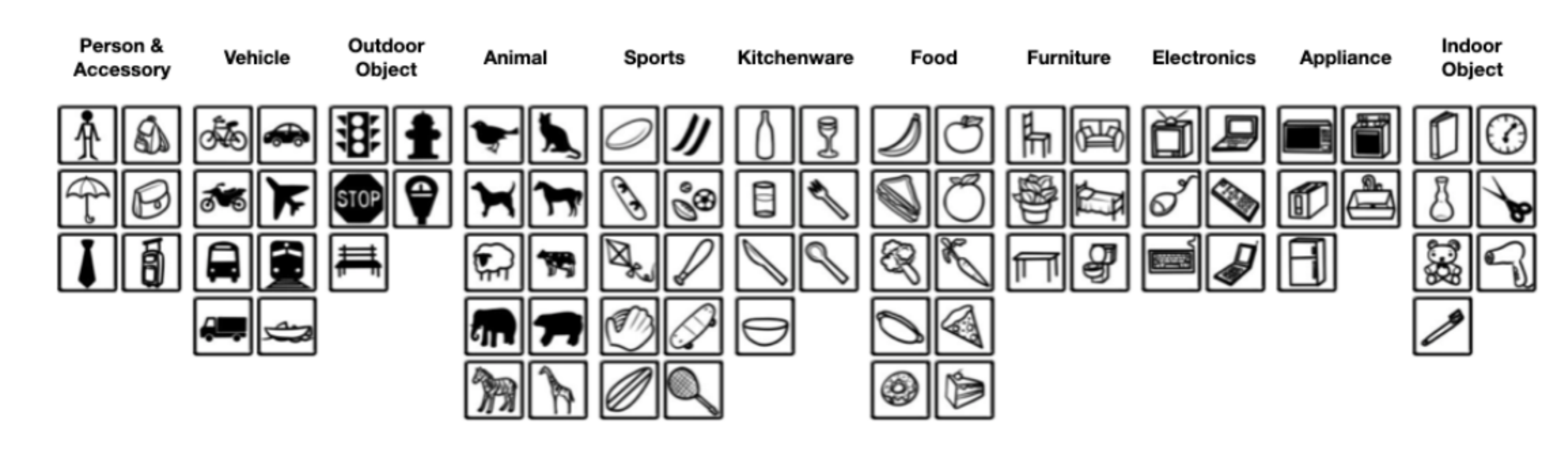}
    \caption{There are 80 object categories classified into 11 super-categories in the MS-COCO dataset~\cite{Lin:ECCV2014}. All object annotations are generated by crowd workers from Amazon Mechanical Turk (AMT). The whole dataset is downloaded from the MSCOCO website (https://cocodataset.org).}
    \label{fig:mscoco-data}
\end{figure}

In this work, we choose MS-COCO as our testing dataset based on three major reasons. First, the MS-COCO is one of the most notable benchmarking datasets for object detection, scene understanding, and visual reasoning. Second, the 80 object categories are selected by experts with thorough considerations~\cite{Lin:ECCV2014}. The object categories are specific, and all of them are entry-level categories, i.e., the category labels are commonly used by humans when describing objects. Also, the categories are from a representative set of all categories, which are relevant to practical applications and occur with high enough frequency. Every image captures everyday objects in diverse contexts. Third, every object category contains large amounts of data. The average number of objects per category is 27,472.5, which is the richest dataset for objects in the context. In this experiment, we used the MS-COCO 2017 dataset. There are 123,287 images with 896,782 annotations and 80 object categories. The 80 object categories are grouped by 11 super-categories, including Person \& Accessory, Vehicle, Outdoor Object, Animal, Sports, Kitchenware, Food, Furniture, Electronics, Appliance, and Indoor Object (see Figure~\ref{fig:mscoco-data}).

\subsection{Interactive Interface for Thing Constellation}
\begin{figure}[h!]
    \includegraphics[width=\textwidth]{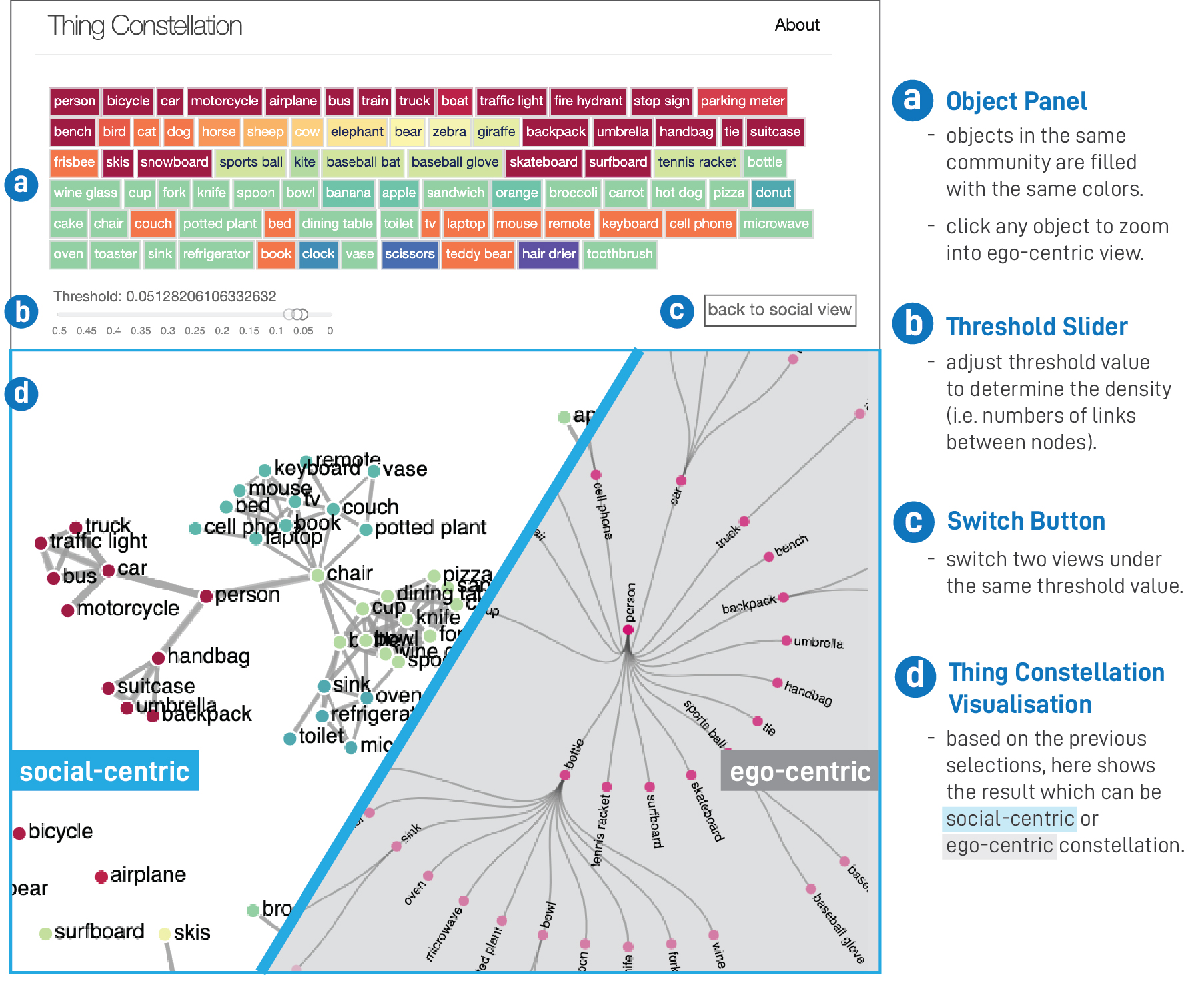}
    \caption{The thingCV consists of (a) object panel, (b) threshold slider, (c) switch view button, and (d) two types of Thing Constellation visualization (i.e., social-centric view and ego-centric view). The tool and source code are released to the public (https://thingconstellation.github.io)}
    \label{fig:thing-constellation-tool}
\end{figure}

We design Thing Constellation Visualiser (thingCV), a tool that visualizes a constellation intertwined with 80 common objects (see Figure~\ref{fig:thing-constellation-tool}). With thingCV, designers can use the threshold slider to observe the changing patterns of the two types of constellations (e.g., social-centric and ego-centric views). By clicking the switch button, the tool will zoom in to an ego-centric view or zoom out to a social-centric view. The social-centric view provides an overview structure of the constellation. Objects with higher scores in co-occurrence will be grouped into the same community highlighted in the same color. Second, the ego-centric view provides a detailed look at an ego-centric constellation. By clicking any objects on the panel, users can jump into different types of ego-centric constellations to investigate specific objects.

\subsubsection{Object panel}
The object panel contains a list of objects in the Thing Constellation. Each object is filled with one color from the spectral diverging color scheme; objects in the same community, calculated by the community detection algorithm, are filled with the same color. Users can click an object to trigger an ego-centric constellation with a focus node (i.e., the clicked object).

\subsubsection{Threshold slider}
Thing Constellation is a weighted social network of objects where the links among nodes (i.e., objects) have weights (i.e., co-occurrence scores) assigned to them. The threshold slider is designed to filter links that are smaller than a threshold value. Users can use the slider to adjust the threshold value to determine the density (i.e., the number of links between nodes) of the social network. The threshold value is ranging from 0.5 to 0, which is determined by the actual weight distributions calculated from the dataset. The constellation with the highest threshold value (0.5) represents a fully disconnected network; by contrast, the constellation with the lowest value (0) represents the original aspect of a social network with all links (i.e., 2686 links). The community detection is triggered immediately while changing the threshold value, and all objects will be assigned to their belonged communities. The objects in the same community are filled in the same color. Through the threshold adjustment, users can see a dynamic change of network structure and changing communities---groups emerge or disappear, multiple groups merge together or break into small pieces.

\subsubsection{Switch button}
The switch button is used to change between the social-centric and the ego-centric view under the same threshold value. With this feature, users can flexibly decide which aspect they want to focus on and use the slider to see the changing patterns. Also, the interface only presents one type of constellation at the same time with the goal of allowing users to focus on one perspective for deep investigation.

\subsubsection{Social-centric view}
The social-centric view presents a social-centric social network using a force-directed graph layout~\cite{Bannister:GD2012}, where nodes within the same community are colored in the same color. Each node is clickable and users can click a node to enter an ego-centric view.

\subsubsection{Ego-centric view}
The ego-centric view presents an ego-centrical social network using a Tidy Tree Layout~\cite{Bostock:IEEE2011}. The network consists of a central node as the target object, and its neighbor nodes are presented in outside circles. The neighbor object is shown at the different levels of circles based on the distance between the target object and themselves. The ego-centric network is expanded by the spreading activation technique with a decay factor of 0.8 and a firing threshold of 0.05. The values (e.g., decay and threshold) are determined by iterative experiments. 

\subsection{Implementation}
To build a social network of objects, we used COCO API to retrieve all images based on a specific category and calculate a co-occurrence score between every two objects based on our proposed method. In the end, we built a social-centric social network based on 123,287 images. The network has 80 nodes and 2686 edges; the average degree of the graph is 67.15. Each edge has its own weight, ranging from 0 to 1. We built a web-based visualization tool that uses jLouvain.js\footnote{https://github.com/upphiminn/jLouvain} to detect object communities and uses D3.js\footnote{https://d3js.org} to visualize the constructed network for allowing design researchers to explore complex relationships among everyday objects through global (i.e., social-centric constellations) and local perspective (i.e., ego-centric constellations). The interactive tool and source code are both released to the public\footnote{https://thingconstellation.github.io}. 

\section{Design Workshops for Exploring Thing Constellation}
\label{sec:design-workshops}
We conducted two workshops with a total of 8 participants (4 males and 4 females) (see Table~\ref{tb:workshop-participants}). All participants are designers (half from industry and half from academia) and have professional experiences in IoT design or product design. The goal of the workshops is to explore how designers use our tool to make sense of co-occurrence relations among everyday objects and investigate whether visualizing connected objects can be used to explore future IoT design. Each workshop was 90 minutes long and took place in Taiwan. 

\begin{table}[]
\begin{tabular}{|l|l|l|l|l|p{5.4cm}|}
\hline
& \textbf{Participant} & \textbf{Sex} & \textbf{Age} & \textbf{Professional Role} & \textbf{Professional Experience}  \\ \hline
\multirow{4}{*}{W1} & P1 & M  & 31-35 & Product Designer & Two years working experience on designing smart lock systems \\ \cline{2-6} 
& P2 & F & 31-35 & UI Designer & One year working experience on intelligent lighting patterns for architecture  \\ \cline{2-6} 
& P3 & M & 31-35 & Tech Lead & Three years working experience on developing IoT systems with the integration of voice agents for hotels \\ \cline{2-6} 
& P4 & M & 31-35 & Exhibition Designer & Three years working experience on IoT-related design research, and applying IoT into designing interactive exhibitions \\ \hline \hline
\multirow{4}{*}{W2} & P5 & F & 26-30 & IxD PhD Student & Three months working experience in a technology company, but limited experiences on IoT\\ \cline{2-6} 
& P6 & F & 31-35 & IxD Master Student & Eight years work experience on media design before doing academic research. Current research focuses on designing IoT interaction with cats. \\ \cline{2-6} 
& P7 & F & 26-30 & IxD PhD Student & Major research is design phenomenology. Current research is designing intelligent objects. \\ \cline{2-6} 
& P8 & M & 26-30 & IxD Master Student  & Study on industrial design, electrical engineering, and Photography. Current research is designing intelligent objects. \\ \hline
\end{tabular}
\caption{Eight participants in the two exploratory workshops.}
\label{tb:workshop-participants}
\vspace{-3em}
\end{table}

\subsection{Workshop Procedure}
\begin{figure}[h!]
    \includegraphics[width=\textwidth]{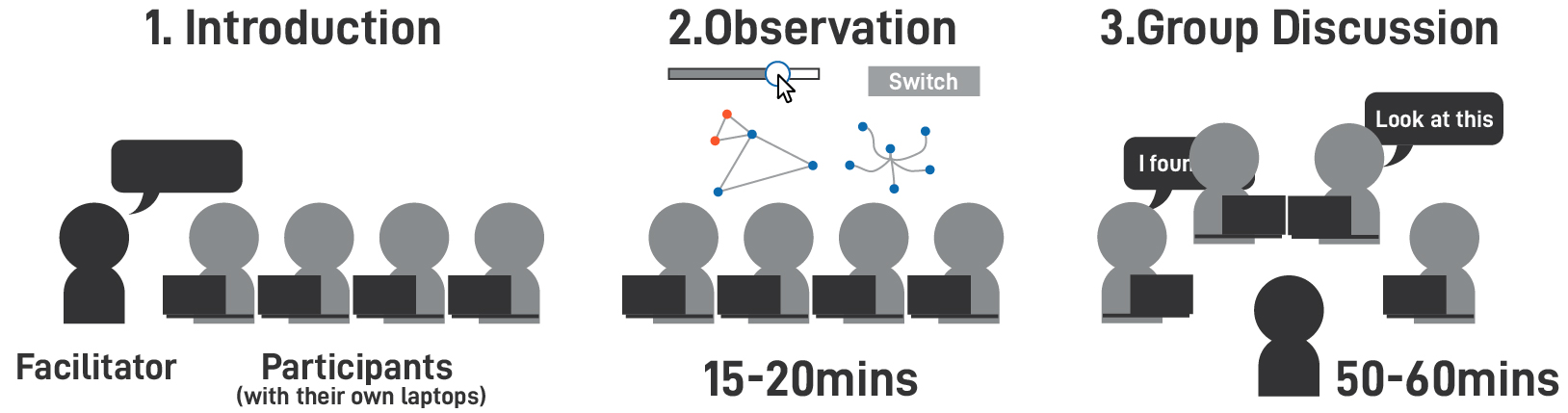}
    \caption{The workshop consists of three sections. 1) Introduction: the facilitator (i.e., one of the authors) introduces the tool and task. 2) Observation: the facilitator asks participants to play with the tool and capture the interesting screenshot of the tool. 3) Discussion: the facilitator hosts a discussion with participants to share their findings and reflections.}
    \label{fig:workshop}
\end{figure}

The workshop was composed of three sections: 1) introduction, 2) observation through the tool, and 3) group discussion (see Figure~\ref{fig:workshop}). In the introduction section, the researchers first welcomed every voluntary participant to join this workshop and asked them to introduce themselves. Then we introduced the purpose of this workshop---inviting participants to test and play our developing tool, thingCV. We introduced the tool, which provides designers a new way of investigating their everyday objects from the field for IoT design inspirations. We also explained that the tool is a data-driven visualization tool to present the co-occurrences of things (e.g. living and nonliving objects) from millions of online photos shared by the general public. Then, the researchers demonstrated the functionalities of the tool on the laptop, such as a threshold slider to adjust the number of links in the network, a switch button to change two different thing constellations (i.e., social-centric and object-centric view), and a hotkey to search for specific objects. After the introduction, the participants were given a link to access the thingCV and used the tool on their own laptops (they were told to prepare their own laptops before the workshop). 15 to 20 minutes were given in this observation section. During the observation section, we asked participants to play with the tool and captured any interesting findings by screenshots or took notes on the paper. Finally, one of the researchers, as a facilitator, hosted a group discussion with participants. 

During the group discussion, three topics were generally given by the facilitator to guide participants to discuss one by one. First, the participants were asked to share their first impressions and experiences of using the thingCV individually. Second, the participants took turns to present their interesting findings and discussed them with other participants. In the end, all participants were asked to summarise their most interesting objects during their discussions. The participants were also asked to reflect on these findings and share their new understanding or ideas inspired by the tool. Note that the participants did not always follow the topics given by the facilitators. Instead, the facilitator even encouraged the participants to actively share, debate, or speculate more on other topics related to the tool with other participants. The facilitator only controlled the time and sometimes cued the participants to make sure each of them responded to the main three topics in an equal time. Finally, two design workshops were successfully conducted, and we gathered rich feedback on the tool and findings with eight experienced designers. The total length of the workshop was 90 minutes, and all of the discussions were audio-recorded, transcribed, reviewed, and summarized into transcripts for data analysis. 

\subsection{Data Gathering and Analysis}
After the workshops, we analyzed the collected data, including a transcript of the audio recording of the workshop, screenshots captured by participants, and paper drawings and notes taken by the participants during the observation with the tool. Researchers used thematic analysis~\cite{Braun_Clarke:2006} to analyze the collected data collaboratively to obtain four main themes with related findings.
{\color{change}The coding and analytical procedure includes three stages. First, two researchers (i.e., two of the authors) read through all the interview data, including audio transcripts, screenshots, paper drawing and observation notes; then, they identified the common features in the data for further analysis. The two researchers annotated and discussed the data scaffolded by the group discussion structure of the workshop, which are tool usage process and experiences (i.e., how they use and feel about the tool), data interpretation (i.e., what they found in the tool), and general reflection (i.e., what new understanding they obtained while using the tool). Second, the researchers prepared and presented all annotated data to a group of experts who have mixed backgrounds in design and computer science (i.e., all authors). The researchers carefully examined the data, compared the initial themes with raw data, and developed sub-themes. Last, the researchers (i.e., all authors) combined and refined sub-themes in an iterative, dialogic process until everyone agreed and then the final themes were generated. The final themes synthesize how thingCV engages participants to investigate the emergent relationships among everyday objects.}

\section{Findings}
\label{sec:findings}
The qualitative data analysis resulted in four themes: 1) observing thing constellation flexibly, 2) projecting social quality onto things, 3) discovering emerging diverse contexts via object clusters, and 4) changing their perspectives to revisit everyday practice. 

\subsection{Observing Thing Constellation Flexibly}
The results showed that all participants were engaged in making sense of thing constellation with our tool through a flexible exploratory process. They have discovered their own interesting objects and gained new understandings of everyday social practices by revisiting familiar objects from an alternative perspective. The participants' first impressions of using the tool are very positive. For example, ``\textit{This is so cool.} (P1)'', ``\textit{Cool! I like the dynamic effect.} (P5)'', ``\textit{I don’t know they can be shown at the same time that often. Amazing.} (P3)'' To further present the benefits of this tool, we showed how the three key features of the tool support the design exploratory process. 

\subsubsection{Perspective-switching button allows zoom in \& out of thing constellation}
The two perspectives (social-centric and ego-centric) enabled the participants to zoom in and out to investigate different angles of the thing constellation. Social-centric perspective allowed them to `zoom out' on the network to interpret how objects were connected with each other, such as ``\textit{I firstly use the social-centric perspective to observe an overview of the network and identify how social and lonely an object can be.}'' (P2). Whereas the ego-centric perspective allowed them to `zoom in' the network to look into the details of the connections to a specific object: ``\textit{then I use the object-centric perspective to look into the details of how objects link to each other.}'' (P2). By switching back and forth between two perspectives, participants not only looked into the details to unpack the social life from a single object perspective but also examine the entire social network intertwined by various objects.  

\subsubsection{Abstract network representations promote open interpretations on thing constellation}
While the object instances presented by our tool were collective objects extracted from millions of everyday photos, the participants were able to use such an abstract form of a social network of collective objects to discuss interesting findings with other participants within the group. They have speculated various potential scenarios and gained new understandings of everyday objects. Such abstraction creates an ambiguity of `what these objects are exactly', which sparks active discussions among participants. For example, `person' was interpreted differently by participants. P5 recognized `person' as users to interact with diverse types of objects; however, P8 interpreted `person' as non-users such as pedestrians or other family members because ``\textit{`person' can be recognized because they are non-users who is taken photos by `the user'---the one who controls the camera.}'' Taking ``bear'' as another example, participants interpreted ``bear'' as different instances, such as `the real bear living in the zoo (P5)', `the toy bear sitting on the bed (P6)' or `the bear from the painting (P8)' so that participants speculated diverse contexts and had active discussions with others. Every participant had their own interpretations of these abstract nodes or links. Therefore, the tool enables a platform for every participant to share, discuss, and even debate with each other. In a way, the ambiguity of such data representation opens an open space for every participant to explore potential possibilities happening in the thing constellation. 

\subsubsection{Interactive threshold enables dynamic constellation observations}
Participants were engaged in observing the dynamic changes between nodes and links by using the interactive threshold. For example, ``\textit{I am pretty engaged in seeing their dynamic changes. It is very cool to see which object is the latest one that `joins' the group.} (P1)'' By adjusting the threshold, participants were able to explore and identify their own interesting segments of the constellation. While they used the same tool with data visualization, participants discovered very different things in objects and object communities. As a result, such diversity sparked interesting discussions among participants. Participants were inspired by others and interpreted more diverse contexts and developed new understandings by revisiting their everyday practice with objects. For example, P5 initially pointed out `book' and further shared his/her interpretation that ``\textit{`chair' is the key object to connect `book' and `person'.}'' Inspired by P5, P7 took a further look at the community which `book' was in and found other objects different from the `chair': ``\textit{Really? That is interesting! Let me check, too (...) but I found that `bottle' is the real key object to connect `book' and `person'.}'' Furthermore, P7 found more possible links around the book. While both of them observe the book with different thresholds, they were able to see the distinct structure of the network and were engaged in discussing various contexts for the target object (i.e., book). For example, P5 speculated a context where the book is the one used to lift the chair. Then, P7 shared another context where an IoT system asks the bottle to look for the book. To sum up, the interactive threshold allows participants to explore their own interesting segments and discover more contexts with other people.

\subsection{Projecting Social Quality onto Things}
During the workshop, we found that participants projected their social experiences onto things. When they observed the dynamic changes of thing constellation, they used ``social roles'' in people's networks to describe the social characteristic of identified objects and groups of objects. For example, the participants have identified levels of sociability for each object based on the number of social links for each object. The object connecting to more other objects is a sociable object, which indicates that they have many friends; the object connecting to no object is an isolated object, which indicates that they do not have friends. Also, the participants looked into the network structure and identified the direct and indirect links for a group of objects. They interpreted the objects having different social distances and further identified bridging objects which connect to two different objects or clusters. Finally, the participants observed a dynamic sequence in which objects connected to each other at different points in time. They interpreted this sequence as joining time for each object: ``\textit{I firstly noticed `chair' is the first one that `joins' the group.}(P6)'' The following examples showed how the participants identified different social levels of an object from sociable to isolated and different joining time from active to passive.  

\subsubsection{Person and couch are sociable and busy.}
\begin{figure}[h!]
    \includegraphics[width=\textwidth]{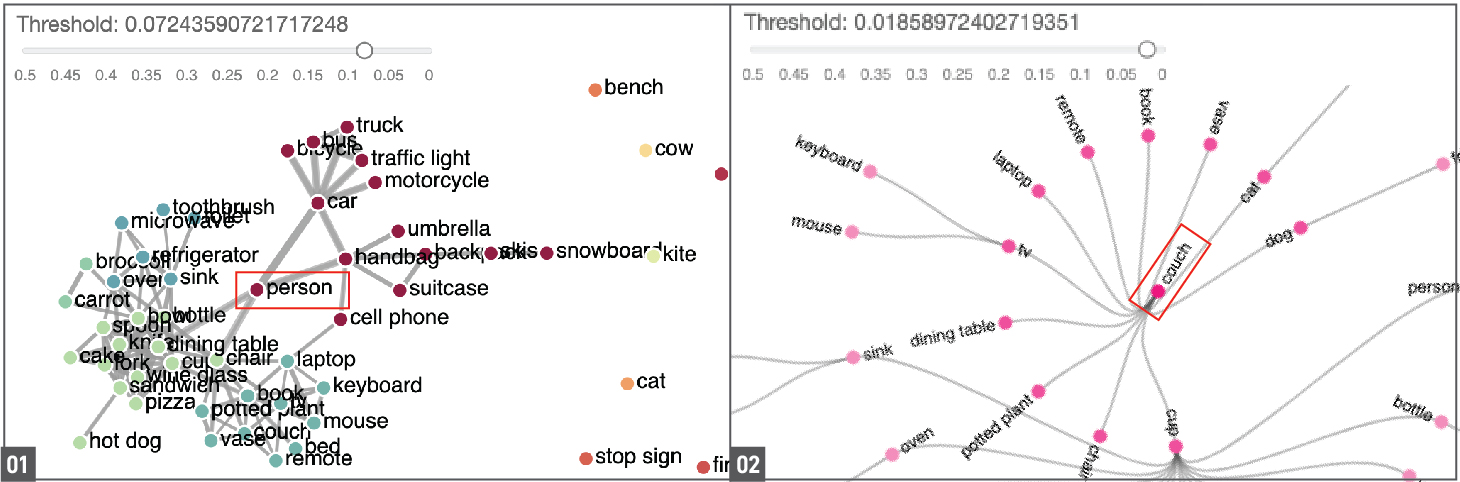}
    \caption{The sociable object is the most connected objects in the network. Two selected examples shows 1) Person (social-centric view), and 2) Couch (ego-centric view) are sociable.}
    \label{fig:sociable}
\end{figure}

Participants described the objects that are linked to the most as the sociable and busy objects such as `person' and `couch' (P7) (see Figure~\ref{fig:sociable}-1 \& 2). These objects are linked with multiple object clusters. Participants saw such an object as a sociable object which is involved in various social activities. As such, P7 was inspired by these sociable objects and found a new design opportunity in the future IoT design. For example, P7 shared that ``\textit{couch is connected with various objects to support diverse activities in people’s everyday lives. Maybe the couch can be the perfect IoT object or the interface to control various activities or do something smart.}'' However, not all of the participants were interested in sociable objects. For example, P4 was less interested in identifying sociable objects because ``\textit{these objects can work and be placed in any context. It is a bit difficult for me to be immediately inspired by any special context.}'' Nevertheless, sociable objects were still discussed within the group, and played an important role for many participants to compare with the opposite ones, isolated objects. 

\subsubsection{Hair dryer is the most lonely object.}
\begin{figure}[h!]
    \includegraphics[width=\textwidth]{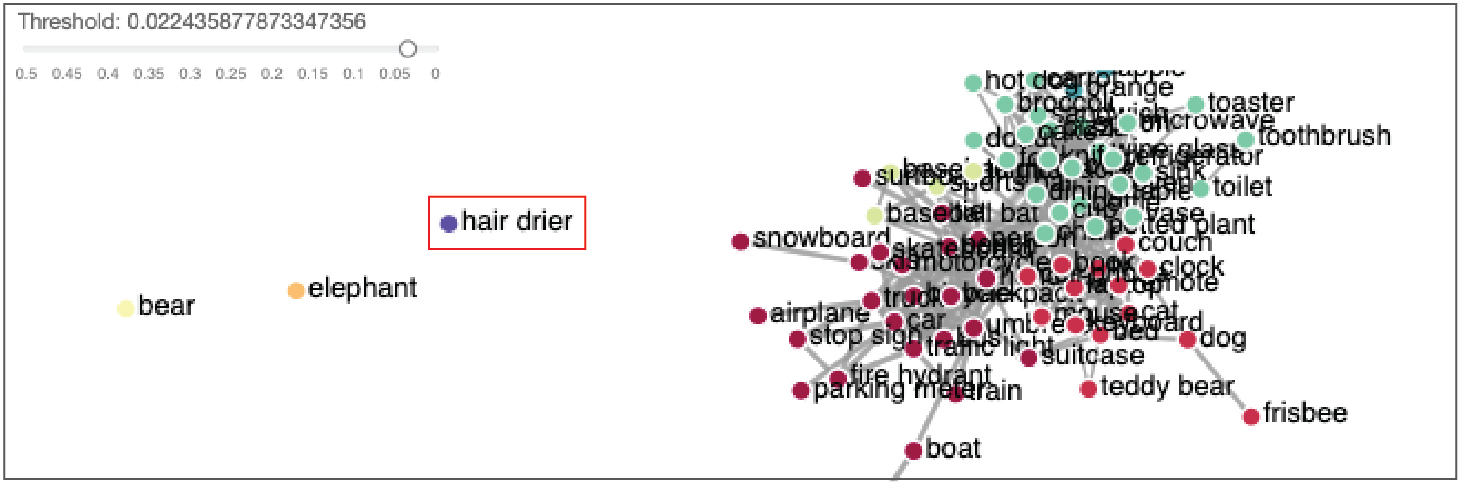}
    \caption{The isolated object is the least connected objects in the network. One selected example shows `hair drier' is floating alone (social-centric view).  }
    \label{fig:isolated}
\end{figure}

Participants were interested in isolated objects, which have no links with other objects. `Hairdryer’ is one of the most discussed isolated objects in the two groups. When adjusting the threshold from high to low, participants (P1, P4, P6, P7) found that the hair dryer was the object which was always floating alone (see Figure~\ref{fig:isolated}). P1 and P4 actively discussed how it could be so lonely because they used this object almost every day. They kept asking why questions; P1 even compared the hair dryer with other unfamiliar objects (e.g., snowboards and animals), which are barely seen in his/her everyday life, to find the possible reasons. P1 was surprised ``\textit{how could the hair dryer be as lonely as animals? The hair dryer is a common object that appears every day but it is more isolated than the snowboard.}'' P7 found that the hair dryer is less active than the elephant. ``\textit{the elephant joins the group earlier than the hair dryer, how comes?}'' At the end of the discussion, P6 felt sorry for the hair dryer which has no friends and said ``\textit{what a poor little thing!}'' Participants further discussed any possible contexts or reasons to make the hair dryer alone. For example, ``\textit{ the hair dryer was usually used alone} (P4)'' or ``\textit{maybe we seldom put them on the surface so that the hair dryer could possibly link with other objects. We stored them in the cabinet immediately after using the hair dryer.} (P8)'' Hair dryer is one of the most common everyday objects which could be seldom discussed by designers; however, by the use of the tool, the hair dryer became the most special object to be discussed.

\subsubsection{Camera is the outlier (invisible tool guy)}
Besides identifying objects presented on the tool, P6 also identified a missing object, a camera, which was not shown on the tool. P6 said, ``\textit{these objects were all taken by cameras, but the camera is not linked to anyone and is even invisible on the network (...) The camera is left behind and working as an invisible tool guy to document others’ lives without his/hers.}'' P6 interpreted that the camera always watched other objects' activities, but it never joined them. This finding also made the other participants continually discuss and agree that the camera is the truly isolated object, the outlier of the social network. 

\begin{figure}[h!]
    \includegraphics[width=\textwidth]{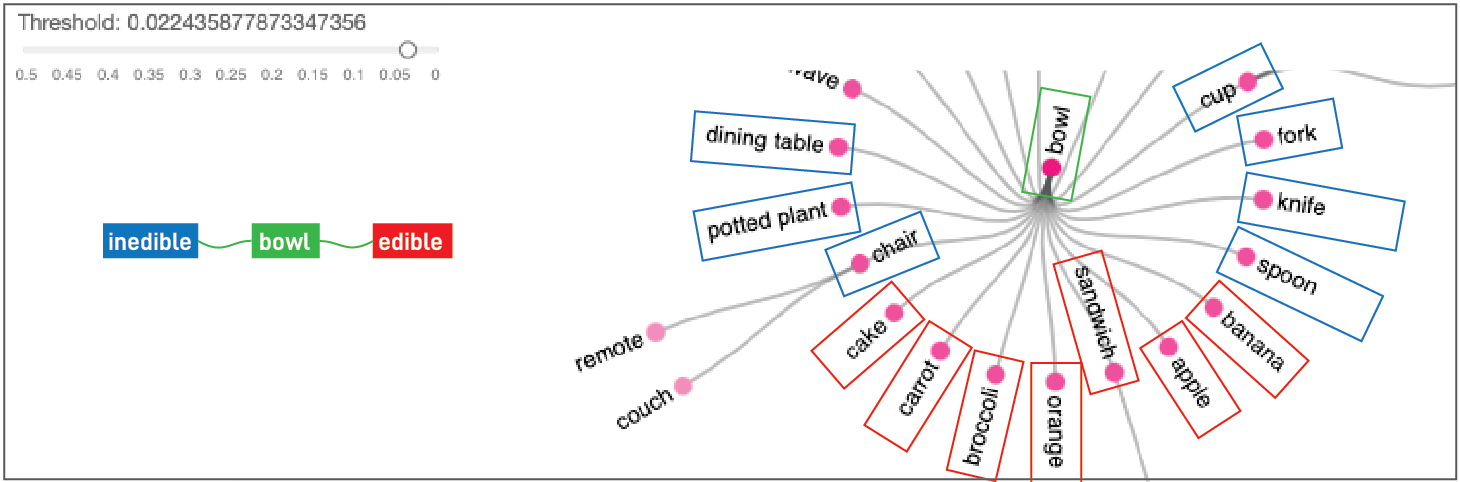}
    \caption{The bridging object is the bridge to connect the two different qualities of object clusters. One selected example shows `bowl' is the bridge to connect the edible (red boxes) and inedible object (blue boxes).}
    \label{fig:bridge}
\end{figure}

\subsubsection{Bowl bridges `food' with the outside (inedible) world} 
P8 found the `bowl' interesting because ``\textit{it bridges two different qualities of objects, edible and inedible.}'' (see Figure~\ref{fig:bridge}) P8 reflected on the everyday practice with food that ``\textit{without bowl, the food can never be exposed in the outside world (socially interact with other types of objects, inedible objects).}'' P8 further took an example of the fruit basket: ``\textit{even the fruit, if they want to be placed on the table, they still need a bowl or a basket, and even need to be placed beautifully.}'' In this case, we see participants project and interpret a ritual for objects reflected from their social norms.

\subsubsection{Cat joins the community earlier than dog.}
\begin{figure}[h!]
    \includegraphics[width=\textwidth]{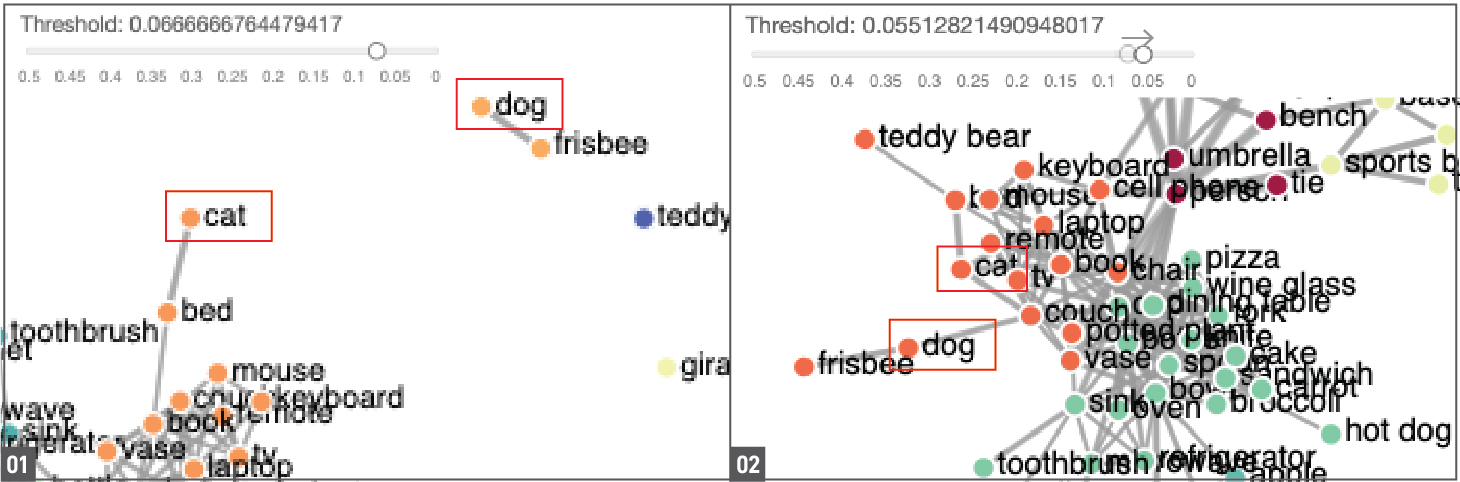}
    \caption{ Participants compared the joining time between two objects, `cat' and `dog': 1) `cat' joins the big community earlier than`dog'; 2) `dog' finally joins the big community later than `cat'.}
    \label{fig:join}
\end{figure}

Participants observed and interpreted different joining times of each object when adjusting the threshold of the thing constellation from high to low. For example, P6 compared two objects, cat and dog (both are usually pets for people), and were surprised by the gap of their joining time: ``\textit{cat joins the cluster earlier than the dog, why!?}'' This finding elicited active discussion among the participants (see two different joining time between cat and dog in~\ref{fig:join}). For example, P5 explained that cats were usually indoor pets and dogs were outdoor pets, so ``\textit{cats can make friends with everyday objects earlier than dogs.}'' Since the tool visualized the photos taken from people, P6 also reflected on the photo-taking preferences from the general public:  ``\textit{perhaps, people like to take photos of cats more than dogs.}''

\subsection{Discovering Emerging Diverse Contexts via Object Clusters} 
\begin{figure}[h!]
    \includegraphics[width=\textwidth]{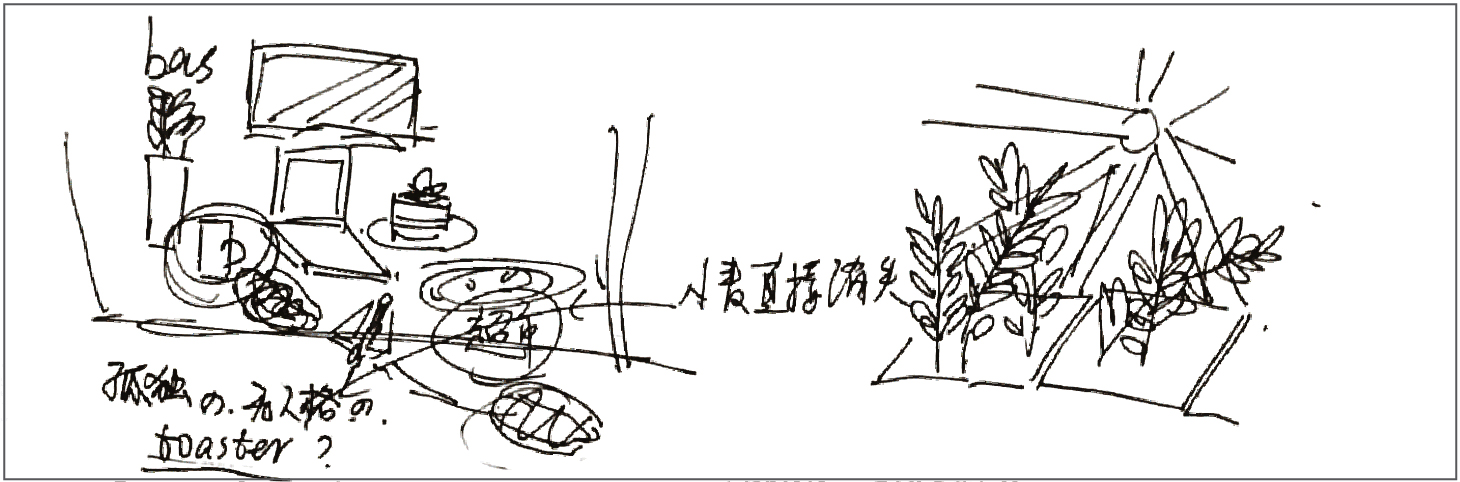}
    \caption{One participant (P5) drew every objects she found in the same network and speculated a possible context around these objects. }
    \label{fig:drawing}
\end{figure}

The desire to map contexts (e.g. scenarios composed of who, where, what, when) onto object clusters (or communities) emerged too:  ``\textit{when seeing `person', `wearables' and `transportations' forming into the same cluster, I immediately have an image about the hustle and bustle city where peoples and cars are crossing by. }(P3)'' Diverse contexts emerged because participants can imagine a space placed by all of these objects from the same cluster, and these objects can support a certain activity altogether. For example, P5 even drew a space situated by these objects from the same cluster: ``\textit{I imagined how they would be placed in the same space. And I found that all of them [the objects] played reasonable roles to support various contexts in the space without feeling out of place.}'' (see Figure~\ref{fig:drawing})

\begin{figure}[h!]
    \includegraphics[width=\textwidth]{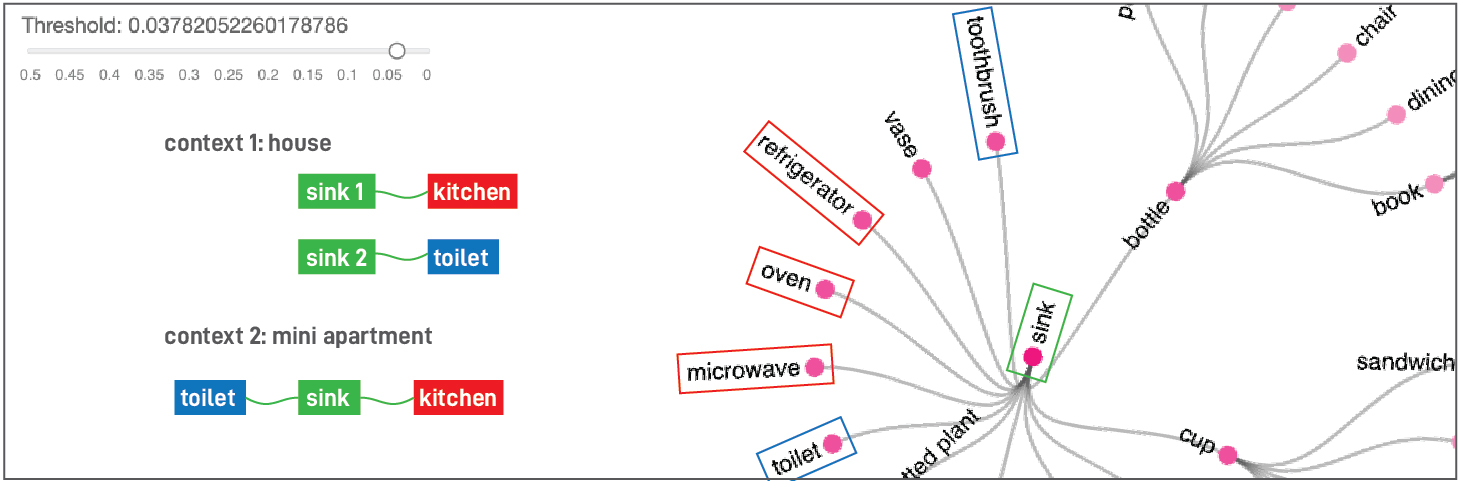}
    \caption{Participants interpreted multiple contexts even if they observed the same object clusters. This example shows two contexts for the sink-related object clusters: 1) people lives in a big house consisted of multiple sinks and each sink support different usages (i.e., washing hands, food, or clothing); 2) people live in a mini apartment with a single sink that is used for all-purpose.}
    \label{fig:sink}
\end{figure}

Moreover, a context changes from one to another by dynamically increasing or decreasing links to different objects with threshold adjustment. For example, P8 interpreted a context transition from a pedestrian walking on the street to a family member interacting at home. Different context mapping can also be found even in the same object cluster. For example, P7 found various contexts for the cluster of the sink: ``\textit{I found different contexts for using the sink such as after using the toilet, cooking in the kitchen, and brushing teeth. These contexts are happening in different spaces. However, it is also possible that this is a mini apartment where all the activities have happened in the same sink.}'' (see Figure~\ref{fig:sink}) Similar context mapping was made by P2 to interpret various contexts for `cup' when observing different links with the cup. 

Finally, participants found that such `emerging contexts' from the tool is useful and can be potentially used for exploring contexts for IoT design. For example, P1 said, ``\textit{without using the tool, I may not be highly aware of so many possible links with a single object. This tool can inspire designers to think about new communications or IoT systems between objects and clusters.}'' Furthermore, the emerging contexts can also be inspiring because participants found that they not only emerge common contexts from familiar everyday settings but also discover or even speculate the unexpected or unfamiliar contexts. Participants even reflected on the missing objects which were not shown on the tool. Thus, the hidden contexts and even the contexts in different cultures emerged. To better present these examples, the following presented their emerging contexts from familiar to unfamiliar, hidden contexts reflected on invisible objects, and contexts in different cultures. 

\subsubsection{Familiar but unnoticed contexts: using a book to lift up the Chair}
A familiar context but usually unnoticed by participants in their daily practice emerged---`using a book to lift up the chair.' When observing the book, P5 was surprised by a social distance between a book and a person: “\textit{why doesn't the book link to the person directly? Instead, there is a chair in-between.}” P5 was curious about this social distance because the first context that came into P5’s mind was a scenario where someone was reading. Such indirect links between book and person inspired P5 to come up with another possible context which is usually unaware of in daily practice: ``\textit{could it be the book used for lifting up the chair by placing it underneath?}'' Sometimes, the chair cannot stand stably due to the uneven rough ground surfaces. P5 found from his/her experiences to project a new context for the book---it is not for reading but for lifting up the chair! 

\subsubsection{Speculative contexts: bottle! please help me call book.} 
\begin{figure}[h!]
    \includegraphics[width=\textwidth]{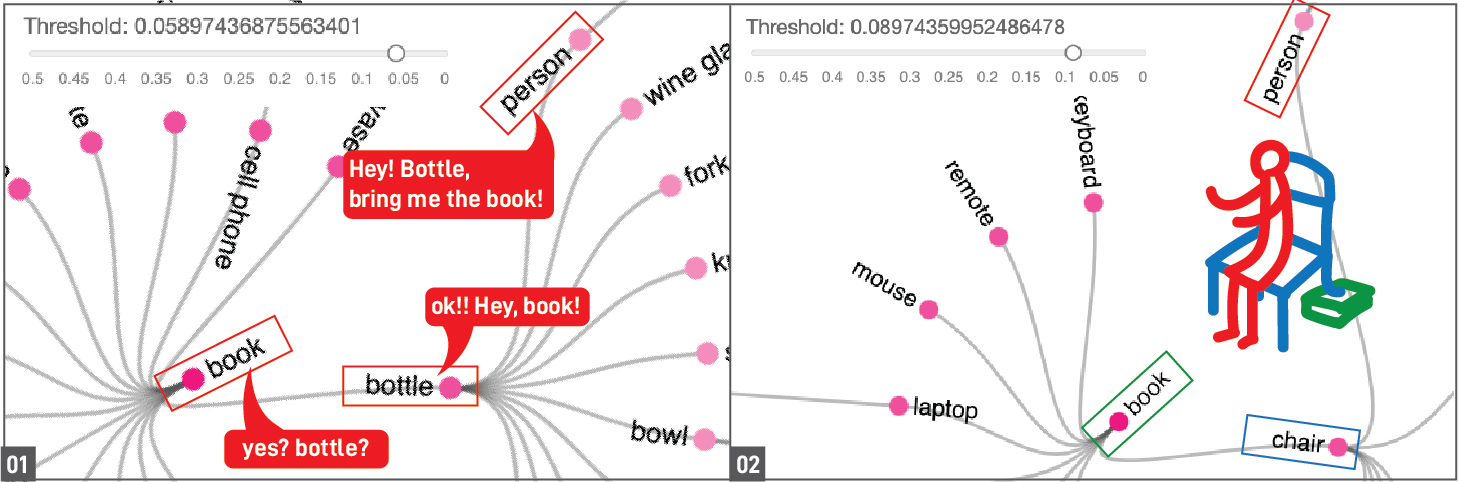}
    \caption{ 1) Participants speculated a context between `bottle', `person' and `book' and imagined a new IoT system for them. 2) Participants reflected on the social links between `person', `chair' and `book', and discovered a familiar but unnoticed context in their prior experiences:  ``the book can be used to lift up the chair.''}
    \label{fig:book}
\end{figure}

A speculative context also emerged when participants identified unfamiliar links between objects such as the link between `bottle', `book' and `person': ``\textit{`person' is indirectly communicating with `book'. `Bottle' is the object which negotiates in-between.}(P7)'' P7 was inspired by this observation and further speculated a fictional scenario and animated objects (see their speculations in Figure~\ref{fig:book}). P7 shared that a bottle that might have an intimate relationship with the book, can be the agent of the book. If someone wants to find any specific book, they can only call the bottle for help. This context seemed to be fictional, but it enabled the participants to identify a ``human-decentralized'' everyday practice, which shows strong evidence that participants changed their perspectives during the observations and explorations.

\subsubsection{Hidden contexts: my intimate cosmetic products}
We also found that participants speculated contexts on those hidden but everyday-used objects which were not shown on the tool. For example, P5 observed that cosmetics and skin-care products were hidden in the social network. P5 further explained that ``\textit{This person doesn't want to let anyone find out he/she will make up. This person wants to make people think he/she is born with beautiful skin naturally. Therefore, these objects (cosmetics and skin-care products) might be secrete and intimate objects which cannot be taken by a camera and shared in public.}'' While the current tool only presented 80 objects, participants are able to speculate more objects and even speculate hidden contexts such as privacy and use preferences.  

\subsubsection{Cultural contexts: who plays with Dog \& where are Chopsticks?}
Contexts in different cultures were also discussed by the participants. For example, P4 found that the cluster of frisbee and dog can be immediately mapped to a context about `people are playing with dogs in the park'. However, P4 found it strange and unfamiliar in his/her everyday practice with dogs. Although this context is quite common in the movie, P4 explained that ``\textit{playing frisbee with dogs is not very common in my country.}'' Instead, P1 added that ``\textit{dogs should be linked with bikes or cars}.'' Additionally, P2 found a different living and eating style than his/hers: ``\textit{the cluster of food (broccoli, carrot, wine), tablewares (fork, knife), snow sports equipment, and home supplements presented a very different lifestyle than mine. The tool presents an exotic western diet and a house with big yards in the snowy north. However, mine is from the eastern culture, which definitely includes `chopsticks' (it is not shown in the tool).}'' When context can emerge by the tool instinctively to participants, these contexts also make participants reflect and discover various cultures which can be depicted by object clusters. 


\subsection{Changing Perspectives to Revisit Everyday Practice}
The participants changed their perspectives to revisit and reflect their everyday practice with objects when switching between the social-centric and ego-centric perspectives. The following presents their changing perspectives elicited by the tool to empathize themselves into different characters and thus to engage with more-than human-centred perspective.   

\subsubsection{Projecting myself into that `object'}
The participants projected themselves as one of the objects to imagine the world they will be living in. For example, P5 shared that ``\textit{I cannot help projecting myself into one of the objects, the `person'. I have noticed any object linked to that `person', and imagine that if I am that `person' what kinds of living I will have, and what objects will contact me.}'' Imagining themselves as being this person switches their perspectives from revisiting their own familiar everyday experiences to engaging within a speculative or fictional context that is shaped by all of the objects.   

\subsubsection{Discovering a human-decentralized everyday}
\begin{figure}[h!]
    \includegraphics[width=\textwidth]{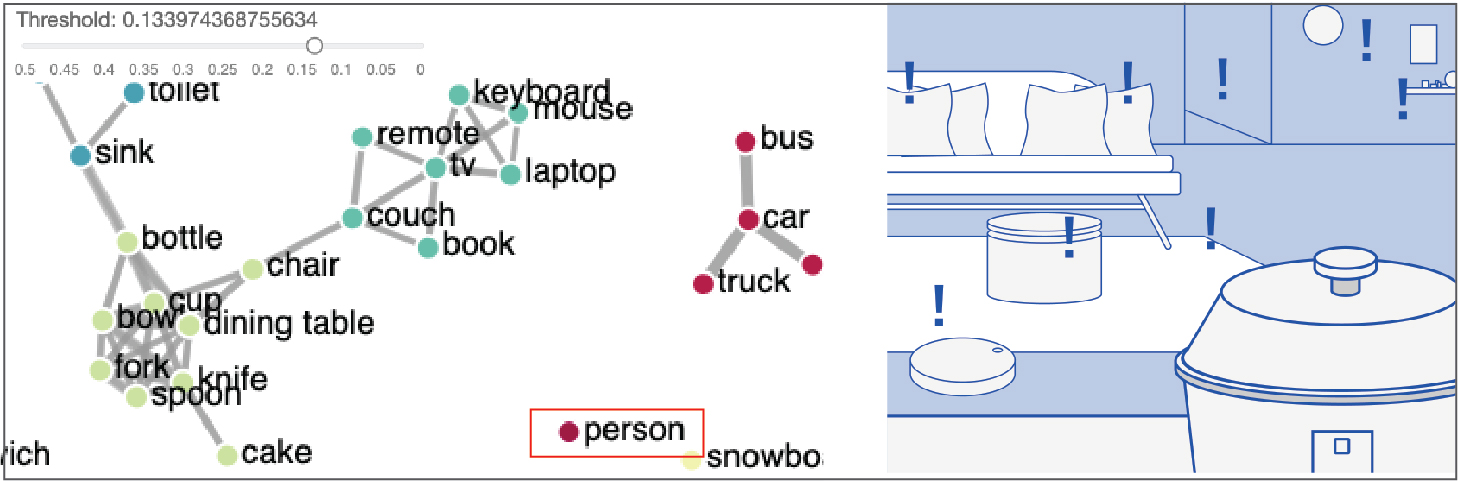}
    \caption{Many objects were connected without being connected to `person'. This picture makes participants change their perspectives in seeing their everyday relationships with things are `human-decentralised' structure. This structure also echos to IoT-related scenarios. For example, the right part shows a scenario that many IoT devices autonomously work and communicate in the background when people is not present.}
    \label{fig:decentralised}
\end{figure}

The participants changed their perspectives from human-centred to more-than human-centred. For example, P7 reflected that ``\textit{before using the tool, I thought every object would be definitely linked to `person', because objects were only used when people touched them based on my intuition. However, it is not a fact at all. During the observation, I found that there are some objects that are already connected as a community, a community without any people involved. And the `person’ is just one of the busy objects like `chair' and `sofa' in the Thing Constellation.}'' P7 was surprised by this finding and described this everyday setting presented by the tool showed him/her a human-decentralized everyday practice. P7 interpreted that ``\textit{It perfectly and truly illustrates the nature of IoT because not every object is fully controlled and surrounded by humans (users). In the background, there will also be some autonomous objects working and communicating on their own without humans.} (P7)'' (see the human decentralized network and the possible IoT scenario in Figure~\ref{fig:decentralised})

\subsubsection{Imagining different constellations to be shaped by a cat}
By playing with our tool, participants were inspired to speculate a different version of Thing Constellation (e.g., a constellation from a cat-centric perspective) by asking a ``what if'' question. For example, P6 speculated, `what if this visualization is based on the data captured by a cat?' P6 said, ``\textit{the current tool only visualizes the photos captured by humans. If these photos are taken by cats, what kinds of Thing Constellation would be shaped? Maybe, the most sociable object would be the `cat food' or `my face' because my cat always slaps my face every morning. }(P6)'' While the current social network of object co-occurrence is analyzed by the photos taken from humans, participants can also be inspired and change their perspectives to speculate a possible Thing Constellation shaped by different species.

\section{Discussion}
\label{sec:discussion}
In the section, we reflect on our exploration process from the beginning of developing the tool to its possible extensions inspired by the workshops with eight designers. First, we discuss object co-occurrence can be a design material to visualize the imperceptible thing constellation. Second and third, we discuss the differences between our work and other relevant research around thing-centred design. {\color{change} Forth, we discuss the role of the underlying dataset in shaping the emergent relationships.} Last but not least, we reflect on the designers' interpretations and see our tool from a visualizer to be a context inspirer. 

\subsection{Object Co-occurrence as Design Materials for Shaping Thing Constellation}
The thing constellation is invisible and can be imperceptible. We regard object co-occurrence as a design material to embody the possible constellation situated in everyday contexts. Object co-occurrence can represent one of our social interactions with objects. To organize our daily routines and activities, people usually arranged the commonly used objects in the same place~\cite{Koren:2003}. These objects, whether to be situated at the backstage or used by us at the front stage, can work all at the same time. For example, watching TV while eating snacks can be represented as a group of front-stage objects that people directly use in the activity, including TV, a remote control, and snacks; at the same time, there are backstage objects that people may not notice during that activity, including tables, curtains, speakers. However, object co-occurrence patterns can capture these objects to represent `the way how people order their lives~\cite{Csikszentmihalyi:1981}'. While co-occurrence is a simplified relation and may not completely represent complex social relations among objects, this work marks a starting point to explore design possibilities of the thing constellation. Our workshops suggested that the designers were able to use this simple form of a co-occurrence object network to revisit their everyday practices and uncover meaningful insights. 

\subsection{A Platform Empowers Designers to Explore Thing Constellation in Their Ways}
Our thingCV makes the thing constellation not only visible but also interactable. Such interactivity (e.g. interactive threshold, zoom in and out in perspectives) empowers a designer to explore the constellation and to identify interesting objects to discuss with the other designers. For example, the interactive threshold allows our participants to adjust and select the constellations that are more meaningful to them. The ability to switch perspectives allows the participants to identify interesting objects and communities and to further speculate on them. Therefore, our interactive interface empowers the designers to interpret the constellation in their own ways, to identify meaningful objects to themselves, and to share their interpretations with others. 

While prior work has investigated the constellation intertwined by various things' perspectives such as the bowl and cup~\cite{Wakkary:DIS2017}, toaster~\cite{Rebaudengo:2012}, and water kettle~\cite{Cila:2015}, these objects are usually pre-selected by the design researchers before the study, whereas there are many more diverse objects in everyday life. As different objects can be meaningful to different people, our tool embraces such diversity and empowers designers to explore the constellation based on their own interests, preferences or needs. Even though there are currently only 80 objects shown in the constellation, this number is not a hard constraint and can be expanded with additional annotations. The objective of developing this tool is to demonstrate the possibility for designers to investigate a relatively large set of things and their constellations. This work thus makes a small but important first step to enable a ``tangible'' constellation for designers to play with exploring these common objects digitally in an interactive way.

This work also echoes the metaphor of constellation that ``different cultures observing the same constellations of stars interpret them variable too.~\cite{Lindley:2020}'' Our thingCV allows every object to have its own place on the stage and invites diverse users of the tool to observe and interpret an object's meanings differently. As Nansen et al. mentioned that every object has its own social life~\cite{Nansen:OzCHI2014}, the thingCV empowers every object to be shown equally as a star, for someone to pass her stories forward. 

\subsection{Abstract Social Network as a Defamiliarized Narrative}
Our thingCV visualizes an abstract social network of the everyday practice which serves as a `defamiliarized narrative'~\cite{Bell:ToCHI2005}. The `defamiliarized narrative' is a common strategy in design research where the goal is to create strangeness to encourage people to revisit their familiar practice from an unfamiliar perspective in order to facilitate their creative decision-making~\cite{Carlson:CC2013}, reflections and speculations to the past, current and future~\cite{Sterling:2009,Sterling:2005}. In our tool, this familiar strangeness comes from the network structure (i.e., nodes and links). Every node represents a familiar object, yet it is also a general and ambiguous entity that is remixed by various living styles, habits, and contexts. As such, the thingCV potentially shows familiar objects but with strange links among them, where these connections can be conflicting to the participants' prior experiences and understanding. For example, some of our participants found a strange network about the `sink'. Several objects which were usually not used in the same context connected to the sink all at the same time. While participants see such network structure was strange, participants were triggered to reflect and speculate its possible contexts. These familiar but strange links stimulate participants' imaginations, open up an interpretation space~\cite{Gaver:CHI2003}, and further lead to discovering new possibilities in their own practices. We see that our thingCV enables such defamiliarized narrative for participants to revisit the everyday relationships from a fresh perspective. 

{\color{change}
\subsection{The Role of the Underlying Dataset}
Dataset plays a pivotal role for designers to dive into the practical world and make sense of what is happening via empirical data. In this work, we chose MS-COCO, an open image dataset, as a starting point to study the relationships between objects and people. By doing so, we can focus on designing rich interactive experiences and interactions for our tool instead of spending too much time on data collection. The selected dataset is a collective dataset in which data are collected by different people and cover diverse contexts. The goal is to understand how designers with no specific interest in a particular context can also resonate with these common objects and social patterns. In this work, we indeed found that the collective dataset can successfully capture familiar and easily understandable relationships known to our participants. The data-driven patterns can facilitate active discussions among people and enable people to compare the detailed nuances with their understandings. Thus, we see our use of the current dataset as a good starting point for researchers to observe and experience a thing constellation that is often imperceptible in their daily contexts.

\subsubsection{Limitations from the underlying dataset}
The chosen dataset would affect the experiences of designers (i.e., tool users) and the insights obtained in the exploration process. First, image data in the current dataset are only collected by people. Different ways of collecting data can provide distinct entry points for people to understand different constellations because data might contain distinct events and social interactions in specific contexts. For instance, our participants speculated about a new constellation shaped by cats' perspectives. Second, our current dataset was collected from the Internet and designed for training a machine to recognize common objects, rather than understanding emergent relationships among objects in the field. Therefore, there might be a gap between the co-occurrence relationships that are detectable in the dataset and those in the real situation. For example, some objects might be occluded, and some objects might be too intimate for people to share online due to privacy concerns. As a result, some relationships may be missing in the current constellation from this dataset. Third, our current dataset consists of images that  present a snapshot of time and do not contain dynamic patterns of the relationships. The way how people organize and interact with objects can change over time, depending on people’s preferences, habits and even emotions. Thus, such a collective dataset might also miss the long-term interaction with an object.

\subsubsection{Possible dataset for capturing emergent relationships}
There is no perfect and fixed dataset for understanding emergent patterns of everyday practice. However, we could articulate what kind of dataset can be used to investigate such relationships between people and everyday objects. First, the dataset can be collected not only by people but also by other non-human actors (i.e., cat, dog, objects). Thus, the dataset can provide richer perspectives about the role of a particular object and its relationships with other objects. Second, any dataset that could be represented as nodes and links can be analyzed by our approach and tool. The node can be a person, an object or a concept; the link can be one type of relationship between two things. The relationship needs to be defined by researchers, such as object co-using frequency, or object ownerships. To enhance the quality of data, some other considerations, such as capturing perspectives and privacy concerns, can be explored in the future. Last but not least, the dataset can be dynamic and evolve over time. To do so, we can allow the users to add, delete and modify the data in the existing dataset and directly modify the graph (i.e., nodes and links) to enrich important information that might be missing during the data collection or analysis process.
}

\subsection{From Constellation Visualizer to IoT Context Inspirer}
Our workshops showed that the designers can identify social things, discover hidden instances and contexts, and change perspectives, which contributes new insights to IoT design.

The social qualities (e.g., sociable, lonely objects) suggest IoT design could consider different social capabilities to things. While researchers have envisioned that the future IoT can enable a new community consisting of many social agents~\cite{Nansen:OzCHI2014}, this does not mean that every object needs to be sociable and connected. In our findings, some objects can be isolated whereas some can be only connected to a specific object (e.g., bridging object). These various social qualities encourage designers to redefine the social links between objects and be aware of the diversity of social things. In addition, the emerging contexts stimulate possible contexts for IoT design. For example, our results showed that the participants, inspired by the unfamiliar social network, actively discussed the possible implications in their own practice. Our tool not only visualizes the constellation but also inspires different contexts. Finally, the changing perspective encourages designers to see every node (e.g., people, food, objects and pet) as equally important. Our results also suggested new design opportunities could also be generated (i.e. designing for cat) through the exploration process. Therefore, our tool makes designers consider other things, beyond human when designing the future IoT.

\section{Design Implications}
\label{sec:implications}
{\color{change}
Our thingCV presents \textit{Computational Thing Ethnography}, which is an alternative thing ethnography that generates a new thing’s perspective by combining data-driven patterns with human interpretations. In this section, we first discuss the design implication around our vision and then discuss how computational thing ethnography contributes to CSCW reserach.
}
\subsection{Towards Computational Thing Ethnography}
While prior design research uses sensor-equipped objects as co-ethnographers to collect data in situ from a thing perspective and then rely on experts or design researchers to make sense of empirical data to understand humans' social practices~\cite{Giaccardi:DIS2016,Chang:DIS2017}, we apply a very different approach to reuse a large-scale public dataset containing millions of photos shared by people voluntarily. Those public photos have rich information about everyday practices with fewer privacy concerns. However, we have admitted that analyzing public data without permission still raise certain ethical concerns. To reduce the ethical concerns, we decide to remove the actual photos and only keep co-occurrence patterns for further investigations and explorations by designers. Then, we only provide an abstract network representation with nodes and links as a defamiliarized structure to stimulate designers to use their own experiences or imaginations to interpret these emerging patterns extracted from a large amount of data. Our tool provides a mask to protect the actual data to be seen, to be interpreted, or to be judged in the design practice. Furthermore, by using a computational approach (e.g. co-occurrence similarity, community detection, spreading activation) to analyzing and visualizing the data, we are able to observe emerging data-driven patterns through a statistical and computational lens.


{\color{change}
Data need human interpretation to be meaningful. While a large collection of empirical data can reveal emergent patterns of everyday practice, data do not speak for themselves and they require humans to give them rich meaning. As Dourish and G\'{o}mez Cruz argue, ``Data must be narrated—put to work in particular contexts, sunk into narratives that give them shape and meaning, and mobilized as part of broader processes of interpretation and meaning-making~\cite{Dourish:2018}.''  In this work, we see the potential benefits of providing designers the freedom to interact with data by using interactive threshold and perspective-changing functionality. Such interactivity enables designers to add their own interpretations upon data and facilitates active discussions within a group. Particularly, they are able to use different snapshots or angles to discuss a specific object or community, and then discover interesting phenomena that are linked to their prior experiences.

As the first step, this work has successfully demonstrated the benefits of combining data-driven patterns with human interpretations. For the next step, we plan to relax certain constraints to allow designers to modify the thing constellation by bringing their own datasets or specifying their relationships of interest (e.g., similarity of context usage or ownership). Designers can use our tool to explore alternative thing constellations in a more flexible way. For example, designers can incrementally add data into the dataset or replace the entire dataset with their own data collected through sensors or by participants from a specific location or context (e.g., home, hospital, train station, factory, school). By doing so, they can observe and discover different objects, object ecosystems and social patterns that emerge in a specific context, rather than general patterns from collective data (as is the case for the current dataset).

Moreover, we plan to explore the interactive dialog between designers and design materials (i.e., data and their social patterns) by enabling designers to directly modify the graph visualization (e.g., add, delete, or modify nodes and links). The algorithms or parameters used in algorithms can also be changed based on users' needs or the characteristics of the data. By opening up for human modification on the graph, designers can bring their own thoughts into shaping the thing constellation and observe the changing patterns under their adjustments. Such interactive dialog could empower designers to try multiple experiments and gain in-depth insights during the exploration process.
}

We argue that our computational thing ethnographic approach is not to replace experts' unique perspectives but to provide great opportunities to allow people with diverse backgrounds to contribute their ideas, experience, and interpretations on the emerging patterns extracted from empirical data. However, we are aware of several challenges of our computational approach. The major one is that data-driven constellations only represent parts of everyday patterns based on the collected dataset. Data are naturally biased, and thus data-driven insights cannot fully represent the actual picture of global constellations among everyday things. Therefore, we believe that data-driven insights should be combined with designerly interpretations. The ultimate goal of the computational thing ethnography is to complement meaningful qualitative insights elicited from the existing ethnography and design research. Particularly, the computational thing ethnography encourages researchers to make data perform themselves and collaborate with data to investigate our everyday practices towards a more-than human-centred understanding.

{\color{contribution}
\subsection{How Computational Thing Ethnography Contributes to CSCW}
CSCW is a research field that is firmly grounded in ethnographic studies of collaborative activities. Blomberg and Karasti reflect on the important role of ethnography in CSCW and suggest an alternative way of repositioning ``ethnography not as a tool for design but as deeply integrated into the doing of design in CSCW~\cite{Blomberg:CSCW2013}.'' In particular, they emphasize that ethnography can contribute to new understandings of the sociality and materiality of work. The goal of this work is to provide a new approach and tool that empowers people to use alternative perspectives to revisit everyday practice and gain an in-depth understanding of emergent relationships among objects. While most ethnographic studies help us see the ``here and now'' and identify temporal and spatial connections among activities, this work contributes a new approach and tool to support designers to explore the possible future by playing with empirical data.

Our work expands the CSCW literature by adding a new concept---computational thing ethnography. We see our tool, thingCV, to be a facilitator to enable designers to actively discuss their experiences and generate insights in collaboration with other people in a group. Our tool also inspires designers to speculate the diverse future IoT scenarios that are grounded in practical data. Moreover, our tool enables designers not only to reflect on their experiences but also to critically compare and contrast the different experiences with others. They can use alternative perspectives to rethink the relationship between people and objects. In this way, it is possible to use our tool to explore cooperative IoT cooperative scenarios.
}

\section{Limitations and Future Work}
\label{sec:limitations-future-work}
We are aware of several limitations of this research. First, our approach only considers object co-occurrence in the same images. In other words, it only captures the degree of frequency of two objects co-occur in the same/nearby location (i.e., ``location/nearby relationship'') rather than other relationships (e.g., usage frequency, utility function, default categories, and other types). Future research is needed to explore different types of relationships among objects and provide insights from different perspectives.
 
Second, we only implement and explore our idea with the MS-COCO dataset, which contains photos mainly taken by Americans. As a result, the constellation visualized by this dataset may only present a limited perspective of people's everyday practice, which is also reflected by our participants (e.g., cultural difference between West and East). However, we are surprised that our participants were not limited by the dataset. Instead, the participants were provoked to actively reflect, criticize and discuss with other people about the difference of their own cultures and Western culture and recognized their stereotypes. Their reflections allow us to see our thingCV is not giving the answer about how people interact with their objects; instead, it provides a data-driven anchor to spark interesting discussions and stimulate people's personal reflections on the social links in everyday things. Nevertheless, future research is needed to explore ways to capture more diverse data and discuss ways to prevent biases in terms of the ethical concerns about the data collection. 

Third, thingCV currently recognizes 80 everyday objects as defined and annotated in the MS-COCO dataset. The proposed data-driven computational framework is general enough to handle larger annotated image datasets with a larger number of objects. In addition, analysis beyond object co-occurrence may be adopted to capture complex relationships among data, algorithms, and other things. For example, to apply our method to support more diverse types of Thing Constellations, we could integrate object detection techniques using the pre-trained model to identify 80 MS-COCO objects or train our own object detector to recognize new types of objects based on newly collected annotations. In addition, we could leverage the power of crowd-sourcing, family-sourcing or expert-sourcing to generate different types of contextual annotations (e.g., object owners, usage time, usage purpose, preferences, etc.) towards a broader perspective.

Finally, we only demonstrate selective showcases and interpretations from our analysis. To get a complete understanding and insights, future research is needed to recruit more people with diverse backgrounds to make sense of these data. In addition, we only investigate how designers can get new understandings of social relationships among common everyday objects through this interactive tool. We have not yet integrated our tool into the actual IoT design process. Our next step is to explore how the thingCV affects the actual design outcomes and designers' perceptions and experience during the IoT constellation design process.

\section{Conclusion}
\label{sec:conclusion}
This paper presents an interactive design tool, Thing Constellation Visualizer, which enables perspective-changing design explorations for empirical understandings of emergent relationships among everyday objects. The proposed computational approach constructs social-centric and ego-centric constellations based on object co-occurrence in a large-scale image collection. Insights drawn from two exploratory workshops suggest that our approach and tool can support designers to identify interesting objects within their corresponding communities and further project social qualities onto them. With the support of thingCV, designers can easily change their perspectives to revisit familiar contexts and generate new insights, which will contribute to the future design of IoT ecosystems.

\begin{acks}
We thank all our participants for investing their time and effort in this project. We also thank Nanyi Bi and Yen-Ling Kuo for their proofreading feedback and the anonymous reviewers for their constructive feedback, which has helped improve this paper. This research was supported in part by the Ministry of Science and Technology of Taiwan (MOST 108-2911-I-011-505, MOST 108-2633-E-002-001), National Taiwan University and Intel Corporation.

\end{acks}

\bibliographystyle{ACM-Reference-Format}
\bibliography{CSCW2021-thingCV}






\end{document}